\documentclass{aastex6}

\usepackage{amsmath}

\begin{document}


\title{A Magnetohydrodynamic Relaxation Method for Non-Force-Free Magnetic Field in Magnetohydrostatic Equilibrium}



\author{Takahiro Miyoshi}
\affil{Graduate School of Science, Hiroshima University,
1-3-1 Kagamiyama, Higashihiroshima 739-8526, Japan}
\email{miyoshi@sci.hiroshima-u.ac.jp}

\author{Kanya Kusano}
\affil{Institute for Space-Earth Environment Research, Nagoya University,
Chikusaku Furo-Cho, Nagoya 464-8601, Japan}

\and

\author{Satoshi Inoue}
\affil{Institute for Space-Earth Environment Research, Nagoya University,
Chikusaku Furo-Cho, Nagoya 464-8601, Japan}


\begin{abstract}

A nonlinear force-free field (NLFFF) extrapolation is widely used
to reconstruct the three-dimensional magnetic field in the solar corona
from the observed photospheric magnetic field.
However, the pressure gradient and gravitational forces are ignored
in the NLFFF model, even though the photospheric and chromospheric
magnetic fields are not in general force-free.
Here we develop a magnetohydrodynamic (MHD) relaxation method that
reconstructs the solar atmospheric (chromospheric and coronal) magnetic field
as a non-force-free magnetic field (NFFF) in magnetohydrostatic equilibrium
where the Lorentz, pressure gradient, and gravitational forces are balanced.
The system of basic equations for the MHD relaxation method is derived,
and mathematical properties of the system are investigated.
A robust numerical solver for the system is constructed
based on the modern high-order shock capturing scheme.
Two-dimensional numerical experiments that include
the pressure gradient and gravitational forces are also demonstrated.

\end{abstract}

\keywords{Sun: chromosphere --- Sun: corona --- Sun: magnetic fields --- Sun: photosphere}



\def\vec#1{\mbox{\boldmath $#1$}}

\section{Introduction} \label{sec:intro}

Information on the three-dimensional structure of the magnetic field
in the solar atmosphere (chromosphere and corona) is essential
to understand and predict various solar phenomena.
However, the atmospheric magnetic field has not been directly measured,
whereas the vector magnetic field, i.e., all three components of the magnetic field,
on the photosphere can be measured with high accuracy.
Therefore, we must reconstruct the solar atmospheric magnetic field
from the two-dimensional distribution of the photospheric vector magnetic field
\citep[e.g., see review by][]{wiegelmann12}.

Since the plasma beta is low in the solar corona \citep{gary01},
the pressure gradient and gravitational forces are negligible
compared to the Lorentz force.
Therefore, a force-free field assumption where the Lorentz force vanishes
may be appropriate for the coronal magnetic field model.
In particular, a nonlinear force-free magnetic field (NLFFF) model,
where a factor of proportionality between the magnetic field and current density vectors
is not constant, has been widely studied.
Since the NLFFF cannot be analytically solved except specific configurations,
various numerical methods for the NLFFF extrapolation such as
the Grad-Rubin method \citep[e.g.,][]{sakurai81,amari06},
the optimization method \citep[e.g.,][]{wheatland00,wiegelmann04},
and the magnetofrictional or magnetohydrodynamic (MHD) relaxation method
\citep[e.g.,][]{mikic94,valori05,inoue14} have been proposed so far.
All these methods well reproduce analytical force-free field models
\citep[e.g.,][]{schrijver06}.
On the other hand,
the vector magnetic fields on the photosphere and in the lower chromosphere
are not in general force-free \citep{metcalf95}
because the plasma beta below the chromosphere is order of unity or more \citep{gary01}.
In order to use observed non-force-free data as the boundary condition
for the NLFFF model, a preprocessing procedure to minimize an integrated force
over the observed region is required \citep{wiegelmann06b,fuhrmann07}.
The preprocessing, however, may change the solution quantitatively or qualitatively.
In fact, comparative studies using solar-like \citep{metcalf08} or
observational dataset \citep{schrijver08,derosa09} pointed out that
the different NLFFF extrapolation methods with/without the preprocessing
produce different magnetic field geometries and different free energies.

The preprocessing corresponds to an adjustive modeling of the photosphere-to-chromosphere
as the bottom boundary of the NLFFF.
In contrast, a magnetohydrostatic (MHS) equilibrium model enables to use
a non-force-free magnetic field on the photosphere directly as the boundary condition
because the magnetic field in MHS equilibrium does not have to be force-free.
Although general solutions for the MHS equilibrium cannot be obtained,
a special class of what are called linear MHS solutions can be solved by an ansatz
that the electric current consists of both a linear force-free current and
horizontal non-force-free current \citep{low91}.
This linear MHS model was applied to a quiet-Sun region \citep{wiegelmann15}
and an active-Sun region \citep{wiegelmann17}.
On the other hand, in order to solve general nonlinear MHS solutions,
it is necessary to develop a more advanced numerical method.
The Grad-Rubin method has been applied to solve the MHS equations
without \citep{gilchrist13} and with the gravitational force \citep{gilchrist16}.
The optimization method has also been extended for the system
without \citep{wiegelmann06a} and with the gravity \citep{zhu18}.
In these methods except the one by \cite{zhu18},
the pressure on the photosphere is imposed as the boundary condition.
The magnetofrictional method was proposed to compute
the three-dimensional equilibrium in a torus geometry
including the pressure gradient force \citep{chodura81}.
Furthermore, the MHD relaxation method that solves
the full compressible MHD equations has been developed \citep{zhu13}.
The validity has been examined in comparison with
an MHD simulation of an emerging flux \citep{zhu13} and
$H_\alpha$ fibrils in the chromosphere \citep{zhu16}.
The density as well as the pressure at the bottom boundary
is fixed by the values on the photosphere in this method.

In this paper, we propose a new MHD relaxation method
for a non-force-free magnetic field (NFFF) in the MHS equilibrium
including the pressure gradient and gravitational forces,
where properly reduced compressible MHD equations is dynamically solved
as a function of a virtual time $t$.
In the previous MHD relaxation method \citep{zhu13,zhu16},
the temperature distribution in the solar atmosphere is calculated
as a result of the full compressible MHD simulation.
On the other hand, in the present method,
the vertical profile of temperature is given beforehand.
Moreover, the previous method requires both the density and pressure values
on the photosphere in addition to the vector magnetic field,
while this method imposes only the vector magnetic field
as the physical boundary condition.
In Section \ref{sec:mhs}, the MHS equilibrium problem to be solved
in this method is prescribed.
The basic equations and their mathematical properties
are addressed in Section \ref{sec:relax}.
In addition, a robust numerical scheme for the equations is presented.
Simple numerical experiments are performed and discussed in Section \ref{sec:exp}.
Finally, in Section \ref{sec:summary}, concluding remarks are presented.

\section{Magnetohydrostatic Equilibrium} \label{sec:mhs}

Consider a dimensionless magnetohydrostatic (MHS) equilibrium:
\begin{equation}
	\left( \nabla \times \vec{B} \right) \times \vec{B} - \nabla p - \rho g \vec{e}_z = 0,
	\label{eq:mhseqs}
\end{equation}
where $\vec{B}$, $p$, $\rho$, $g$, and $\vec{e}_z$ are the magnetic field, pressure, density,
gravitational acceleration, and unit vector in $z$ direction, respectively.
This indicates that the Lorentz force in the MHS equilibrium does not vanish in general
since it can balance both the pressure gradient and gravitational forces.
We subtract a background hydrostatic field, $p_0$ and $\rho_0$, from (\ref{eq:mhseqs}),
and obtain
\begin{equation}
	\left( \nabla \times \vec{B} \right) \times \vec{B}
	- \nabla \tilde{p} - \tilde{\rho} g \vec{e}_z = 0,
	\label{eq:mhsprtb}
\end{equation}
where
\begin{equation}
	- \frac{\partial p_0}{\partial z} - \rho_0 g = 0.
	\label{eq:mhsback}
\end{equation}
The components deviated from the hydrostatic equilibrium are respectively
\begin{equation}
	\tilde{p} = p - p_0 (z) , \quad \tilde{\rho} = \rho - \rho_0 (z),
	\label{eq:prtbpro}
\end{equation}
which are termed ``pressure deviation'' and ``density deviation''.
Note that these deviations are not necessarily small.

Here we assume that the temperature profile in the solar atmosphere
depends only on the height $z$
and coincides with the one determined by the background hydrostatic field as
\begin{equation}
	T(z) = \frac{p_0 (z)}{\rho_0 (z)} = \frac{p}{\rho}.
	\label{eq:temp}
\end{equation}
Hence,
\begin{equation}
	\tilde{\rho} = \frac{\tilde{p}}{T(z)}.
	\label{eq:ro1d}
\end{equation}
Substituting (\ref{eq:ro1d}) into (\ref{eq:mhsprtb}) yields
\begin{equation}
	\left( \nabla \times \vec{B} \right) \times \vec{B}
	- \nabla \tilde{p} - \frac{\tilde{p}}{H(z)} \vec{e}_z = 0,
	\label{eq:mhst1d}
\end{equation}
where the scale height $H(z)$ is defined by
\begin{equation}
	H(z) \equiv \frac{T(z)}{g}.
	\label{eq:scaleh}
\end{equation}
Consequently, the task to be addressed in this paper is to solve
the MHS equilibrium field (\ref{eq:mhst1d}) with the boundary condition as
\begin{equation}
	\vec{B}(x,y,z=0) = \vec{B}_{ph},
	\label{eq:obsmag}
\end{equation}
where $\vec{B}_{ph}$ is the vector magnetic field on the photosphere.
The background fields are not required under the assumption of (\ref{eq:temp}).
Meanwhile, the profile of $H(z)$ is given beforehand in our model.
This is one of the reasonable and efficient approaches
because the three-dimensional temperature distribution in the solar atmosphere
has not been solved consistent with the magnetic field distribution yet.

Note that \cite{gilchrist16} have adopted a similar approach:
The density and pressure are decomposed into the background components
and their deviations, respectively.
The scale height is given beforehand as a function of the position, not only of $z$.
Their approach is more general
since a different scale height can be considered in a quiet or an active region.
However, it is difficult to know the exact scale height in advance
because the three-dimensional distribution of the magnetic field
must also be determined consistently.
Therefore, in the present MHS equilibrium (\ref{eq:mhst1d}),
the scale height is simply given as in (\ref{eq:scaleh}),
as a function of $z$ using typical or average values in the active region.

\section{MHD Relaxation Method for the NFFF} \label{sec:relax}

\subsection{Basic Equations} \label{sec:eqs}

In the relaxation method for the NFFF model, as for the NLFFF model,
the induction equation for the magnetic field is numerically solved
until a quasi-static state is reached.
The basic equations of our method are given as follows:
\begin{equation}
	\frac{\partial \vec{V}}{\partial t}
	= \left( \nabla \times \vec{B} \right) \times \vec{B}
	- \nabla \tilde{p} - \frac{\tilde{p}}{H(z)} \vec{e}_z - \nu \vec{V}
	\label{eq:vnfff}
\end{equation}
\begin{equation}
	\frac{\partial \vec{B}}{\partial t}
	= \nabla \times \left( \vec{V} \times \vec{B}
	- \eta \nabla \times \vec{B} \right)
	\label{eq:bnfff}
\end{equation}
\begin{equation}
	\frac{\partial \tilde{p}}{\partial t} = - a^2 \nabla \cdot \vec{V},
	\label{eq:pnfff}
\end{equation}
where $\vec{V}$, $\nu$, $\eta$, and $a$ are the fluid velocity,
friction coefficient, resistivity, and a ``pseudo-speed of sound'', respectively.
The equation (\ref{eq:vnfff}) indicates that
the fluid is accelerated not only by the Lorentz force,
but also the pressure gradient and gravitational forces,
and decelerated by the friction that increases with the velocity.
The magnetic field is evolved in time according to the induction equation (\ref{eq:bnfff})
including the motion-induced and resistive electric fields.
In addition, the time evolution of the pressure deviation is taken into consideration.
Since a physically correct time evolution of the pressure is not necessary in our method,
we propose a simple evolution equation for the pressure deviation as in (\ref{eq:pnfff}).
The pseudo-speed of sound $a$ is not a physical value but
a numerical one which should be determined by numerical experiments,
and is given by an arbitrary constant in this paper.
Here we insist that $\tilde{p}$ can be negative
as the absolute value of the pressure does not appear in this system of equations.
Finally, we expect that the MHS equilibrium (\ref{eq:mhst1d}) is obtained
since $\vec{V}$ may approach to $0$ due to the friction in (\ref{eq:vnfff}).
Note that the basic equations of our method result in the equations for the NLFFF model
if $\tilde{p}$ is set to $0$.

\subsubsection{Magnetic Field-aligned Dynamics}

We consider here the magnetic field-aligned dynamics in the case that
$H(z) \rightarrow \infty$.
Since the Lorentz force vanishes along the magnetic field,
(\ref{eq:vnfff}) and (\ref{eq:pnfff}) are combined into
\begin{equation}
	\frac{\partial^2 \tilde{p}}{\partial t^2}
	+ \nu \frac{\partial \tilde{p}}{\partial t}
	- a^2 \frac{\partial^2 \tilde{p}}{\partial x_\parallel^2} = 0,
	\label{eq:ptelg}
\end{equation}
or
\begin{equation}
	\frac{\partial^2}{\partial t^2}
	\left( \frac{\partial V_\parallel}{\partial x_\parallel} \right)
	+ \nu \frac{\partial}{\partial t}
	\left( \frac{\partial V_\parallel}{\partial x_\parallel} \right)
	- a^2 \frac{\partial^2}{\partial x_\parallel^2}
	\left( \frac{\partial V_\parallel}{\partial x_\parallel} \right) = 0,
	\label{eq:dvtelg}
\end{equation}
where $x_\parallel$ and $V_\parallel$ show the coordinate and velocity
along a magnetic field line.
The equation (\ref{eq:ptelg}) or (\ref{eq:dvtelg}) is the so-called
telegraph equation which consists of a wave equation and a diffusion equation.
Thus, bumps of $\tilde{p}$ or $\partial V_\parallel / \partial x_\parallel$
propagate almost at the speed $a$ and diffuse almost at $a^2 / \nu$.
Consequently, in the region that the scale height is large,
the pressure deviation evolves so as to be uniform
along the magnetic field line.

\subsubsection{Gravitational Stability}

A thermal-fluid system, in general, can be unstable to thermal convection
if the gravitational effect is taken into consideration.
In order that the solution of the system
from (\ref{eq:vnfff}) to (\ref{eq:pnfff}) converges to the static state,
the system must be gravitationally stable.

We consider here the case that $\vec{B} = 0$.
In this case, the system becomes linear.
If periodicity in all directions and a constant $H$ are assumed for simplicity,
we obtain the following dispersion relation:
\begin{equation}
	\left( \omega + i \nu \right)^2
	\left( \omega^2 + i \nu \omega - a^2 k^2 + i \frac{a^2 k_z}{H} \right)
	= 0,
	\label{eq:disp}
\end{equation}
where $\omega$ and $k$ denote the frequency and wavenumber.
This leads to a sufficient condition for the stability as
\begin{equation}
	\nu > \frac{a}{H}.
	\label{eq:stab}
\end{equation}
See Appendix \ref{sec:disp} for details.
Since a line-tied magnetic field is expected to further stabilize the system,
the condition (\ref{eq:stab}) is
sufficient to stabilize the system with the boundary condition (\ref{eq:obsmag}).

\subsection{Numerical Methods} \label{sec:solver}

A numerical solver for the system from (\ref{eq:vnfff}) to (\ref{eq:pnfff})
may be constructed such that the second and third terms of the right-hand side
of (\ref{eq:vnfff}) and the equation (\ref{eq:pnfff}) are simply discretized
and added to an existing well-optimized code for the NLFFF.
Here, however, we propose numerical methods that are suitable for
our entire system.

We rewrite the basic equations (\ref{eq:vnfff}) to (\ref{eq:pnfff})
in the conservative form as
\begin{equation}
	\frac{\partial \vec{U}}{\partial t} + \nabla \cdot \vec{F} = \vec{S},
	\label{eq:cons}
\end{equation}
\begin{equation}
	\vec{U} = \left[
	\begin{array}{c}
		\vec{V} \\
		\vec{B} \\
		\tilde{p} \\
		\psi \\
	\end{array}
	\right] , \
	\vec{F} = \left[
	\begin{array}{c}
		\left( \tilde{p} + \frac{B^2}{2} \right) \vec{I} - \vec{B} \vec{B} \\
		\vec{V} \vec{B} - \vec{B} \vec{V} + \psi \vec{I} \\
		a^2 \vec{V} \\
		c_h^2 \vec{B}
	\end{array}
	\label{eq:uf}
	\right] ,
\end{equation}
where $\vec{I}$ is the unit tensor, and
\begin{equation}
	\vec{S} = \left[
	\begin{array}{c}
		- \nu \vec{V} \\
		- \nabla \times \left( \eta \nabla \times \vec{B} \right) \\
		0 \\
		- \frac{c_h^2}{c_p^2} \psi
	\end{array}
	\right] .
	\label{eq:src}
\end{equation}
Here we introduce an additional scalar field $\psi$ proposed by \cite{dedner02}
in order to maintain the divergence-free condition for the magnetic field.
The temporal evolution of $\psi$ and $\nabla \cdot \vec{B}$ obeys the telegraph equation,
where propagation and diffusion speeds are controlled by constants $c_h$ and $c_p$.
Thus, $\nabla \cdot \vec{B}$ is kept as small as possible.

We solve the system of conservation laws (\ref{eq:cons}) using the finite volume approach.
Although simple centered-finite differences may be applicable,
a tuned-artificial viscosity must be required for the numerical stability
particularly in using complicated real data.
Therefore, since the system is hyperbolic (see Appendix \ref{sec:eigen}),
we construct an upwind-type scheme
in which an appropriate numerical viscosity is automatically added.
The numerical fluxes for (\ref{eq:cons}) are evaluated as in Appendix \ref{sec:upwind}.
Moreover, in order to achieve high-order accuracy,
third-order MUSCL \citep{koren93} and third-order SSP Runge-Kutta method \citep{gottlieb01}
are applied.
Note that the present solver which is robust can be applied to the NLFFF reconstruction.

\begin{figure}[t]
\centering
\includegraphics[scale=0.6]{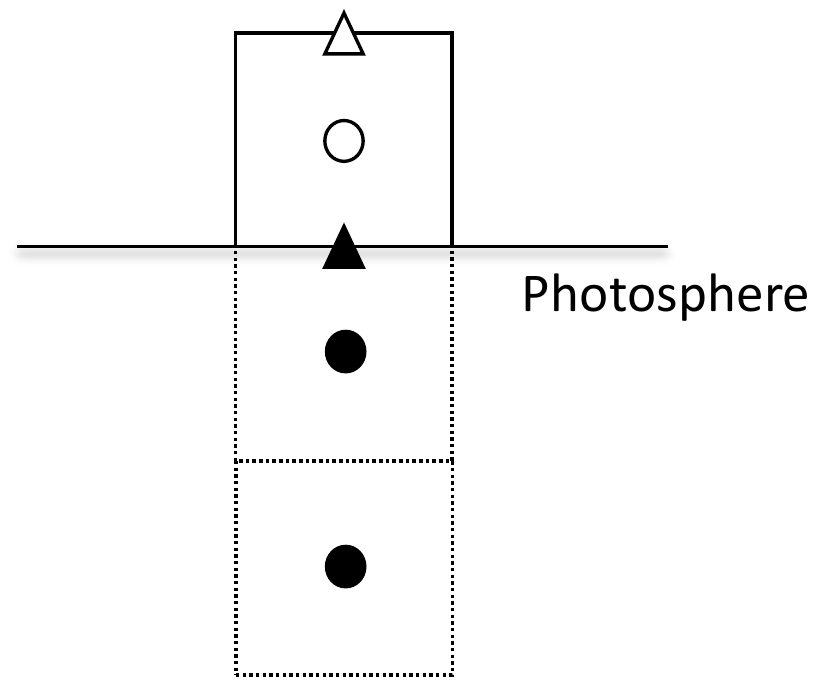}
\caption{
Schematic diagram of the grid structure around the photospheric boundary.
The open circle shows the representative point of a cell
just inside the boundary (termed as $I$),
whereas the filled circles are the points of ghost cells (termed as $G$).
The time evolution of the conservative variables in the cell $I$ is determined
by the numerical fluxes at the open and filled triangles.
The variables or fluxes at the filled marks cannot be treated
in the same manner as those inside the domain.
\label{fig:fig1}}
\end{figure}

We must define the numerical fluxes at the domain boundary
because the boundary lies on a surface of a control volume, i.e., a cell face,
in our approach as shown in Figure~\ref{fig:fig1}.
Since the vector magnetic field on the photosphere is given,
all three components of the magnetic field in the numerical fluxes at the boundary
as $\vec{B}_h$ in (\ref{eq:flux1d}) are fixed by (\ref{eq:obsmag}).
Moreover, two layers of ghost cells,
i.e, the filled circles in Figure~\ref{fig:fig1}, are added outside the domain
in order to compute the other variables for evaluating
the numerical fluxes at the boundary,
i.e., the filled triangle in Figure~\ref{fig:fig1}.
The magnetic field in the ghost cells, $\vec{B}_G$,
below the photosphere is fixed as a function of $x$ and $y$ as in (\ref{eq:obsmag}).
On the other hand, the pressure deviation in the photosphere cannot be measured
with the same degree of accuracy as the magnetic field.
Therefore, the pressure deviation in the ghost cells, $\tilde{p}_G$,
should be given numerically, not observationally.
One approach is that the pressure deviation is determined from
the horizontal force balance in the ghost cells as adopted by \cite{zhu18}.
Another approach is that $\tilde{p}_G$ is extrapolated from the physical values
in the domain.
Here we propose an approximate outgoing characteristics at the bottom boundary,
\begin{equation}
	d w = d \tilde{p}_T- d V_z \sqrt{a^2+B^2},
	\label{eq:apchar}
\end{equation}
where $\tilde{p}_T = \tilde{p}+B^2/2$.
If we set $B = B_{ph}$, ${V_z}_G = 0$, and assume $w_G = w_I$,
where the subscript $I$ is the index
corresponding to the cell just inside the boundary,
we obtain
\begin{equation}
	\tilde{p}_G = \tilde{p}_I - {V_z}_I \sqrt{a^2+B_{ph}^2}.
	\label{eq:bc2}
\end{equation}
Note that positive (negative) ${V_z}_I$ leads to a decrease (increase)
in the pressure deviation.
The top and lateral boundaries should be positioned far enough away
from an active region in general.
Then, for example, we assume a conducting wall, or apply some extrapolation techniques.
The boundary conditions adopted in this paper are summarized in Appendix \ref{sec:boundary}.
Although other extrapolation techniques are worth investigating,
those should be examined using real data and are beyond the scope of present study.

\section{Numerical Experiments} \label{sec:exp}

We demonstrate two-dimensional numerical experiments to investigate
not only numerical stability but also the dependence of $H(z)$
which controls the gravitational effect, in particular.

The vector magnetic field on the photosphere, $\vec{B} (x,z=0)$, is given by
\begin{equation}
	\left\{
	\begin{array}{ll}
		B_x = 0 , B_y = J_0 (x) , B_z = J_1 (x) & \text{for} \ |x| \leq r_b \\
		B_x =  B_y =  B_z = 0 & \text{otherwise}
	\end{array}
	\right.
	\label{eq:bessel}
\end{equation}
unless otherwise stated, where $J_\alpha$ are $\alpha$-th order
Bessel functions of the first kind
and $r_b$ is the first zero point of $J_0$, i.e., $J_0(r_b) = 0$.
The test problems are not so easy from the numerical stability
point of view because $B_z$ on the photosphere is discontinuous at $r_b$.
The pressure deviation at the boundary is calculated by (\ref{eq:bc2}).
The initial magnetic field is given by the potential field
which satisfies the boundary condition of $B_z$.
The other initial conditions are set to zero as
$\left| \vec{V} \right| = \tilde{p} = \psi = 0$.
The dimensions of the computational domain are $-5 \leq x \leq 5$
and $0 \leq z \leq 10$ in the $x$- and $z$-directions, respectively.
The grid numbers are $N_x = 200$ and $N_z = 200$ in each direction.
Number of iterations is fixed to $500000$ in all tests.
The parameters to control the time evolution of the system are given
by specific values as $a = 0.4$ and $\nu = a/H(0)$
except the case $H \rightarrow \infty$ in this paper.
The resistivity $\eta$ is set to $0$.

\subsection{Case $H(z) \rightarrow \infty$}

First, we take notice only of the pressure gradient force.
The gravitational effect in (\ref{eq:vnfff}) is eliminated
if $H(z)$ becomes infinite.
The friction coefficient $\nu$ in this case is given by $0.1 a$.

\begin{figure}[t]
\centering
\plottwo{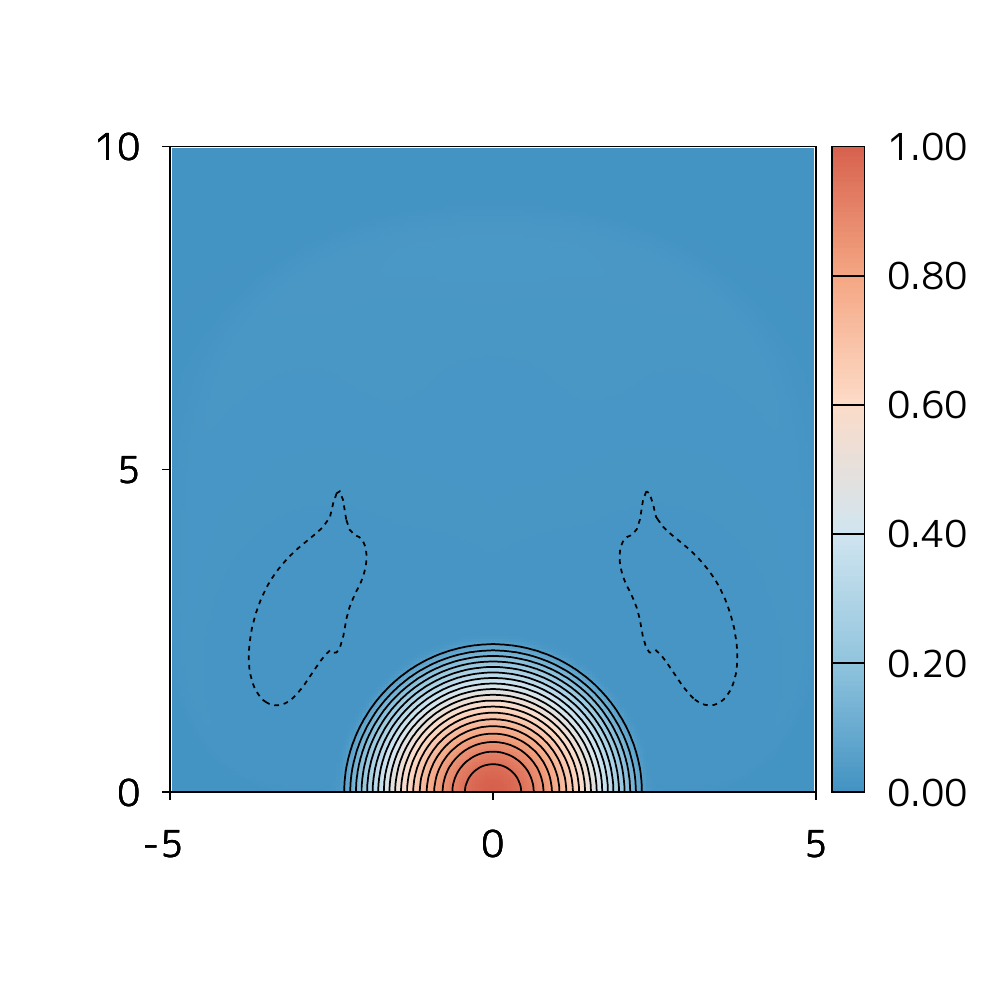}{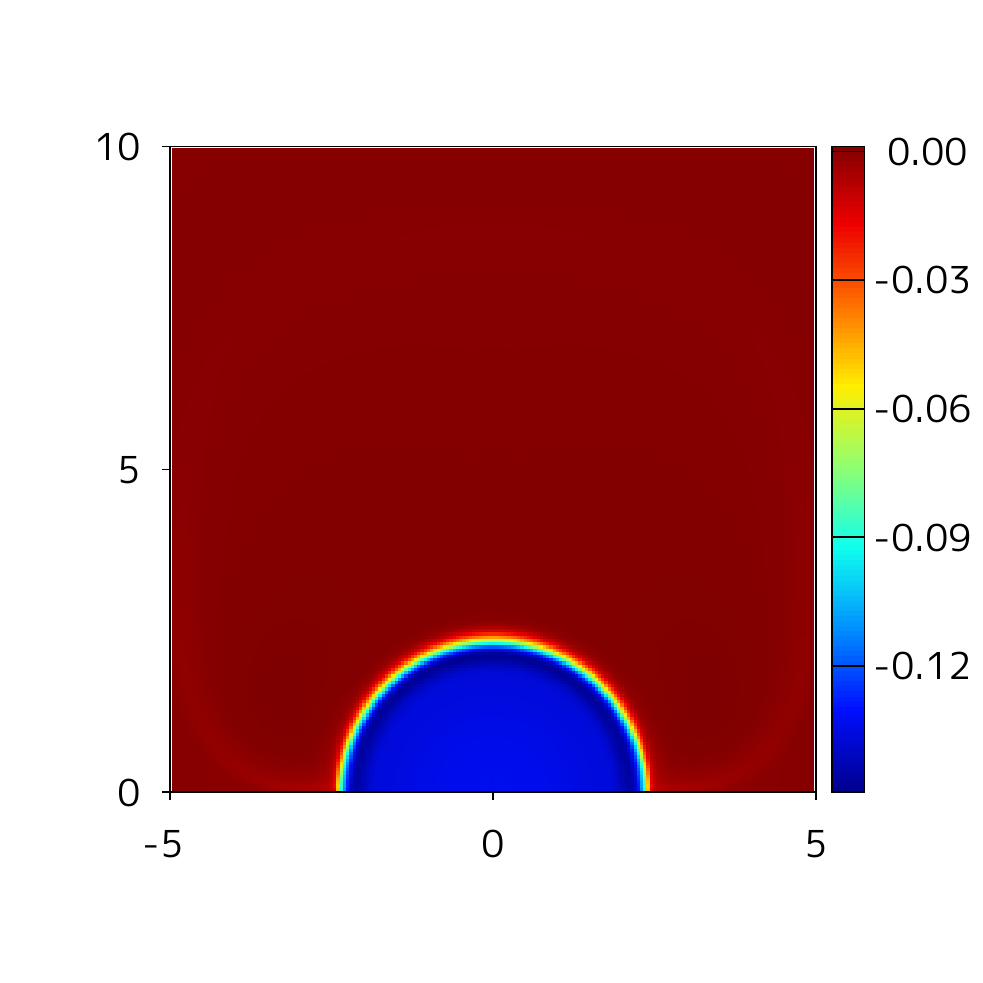}
\caption{
The distribution of the magnetic field and the pressure deviation
for $H(z) \rightarrow \infty$.
(Left) The lines represent contour lines for the magnetic flux-like function
$\Psi(x,z) = - \int_0^x B_z (x',0) dx' + \int_0^z B_x(x,z') dz'$.
When the divergence-free condition of the magnetic field is satisfied,
$\Psi$ completely corresponds to the magnetic flux function.
Thus, the solid and dashed lines indicate the magnetic field lines
connected to and detached from the bottom boundary, respectively.
The color indicates the out-of-plane magnetic field, $B_y$.
(Right) The color indicates the pressure deviation.
\label{fig:fig2}}
\end{figure}

One of the exact solutions is that
a semi-cylindrical force-free magnetic field,
$\left( B_r, B_y, B_\theta \right) = \left( 0, J_0(r), J_1(r) \right)$,
is confined by the surrounding pressure.
Figure~\ref{fig:fig2} shows the distribution of the magnetic field
and the pressure deviation
reconstructed by the present method without the gravity.
The initial potential magnetic field spontaneously develops into
a semi-cylindrical magnetic field in which the pressure deviation is almost constant.
The semi-cylindrical configuration itself is supported by the external pressure.
Note here that the pressure deviation becomes negative inside the magnetic loop
since the pressure gradient force is not dependent on the absolute value of the pressure.
Thus, a Bessel function-like configuration is well reproduced.

\begin{figure}[t]
\centering
\plottwo{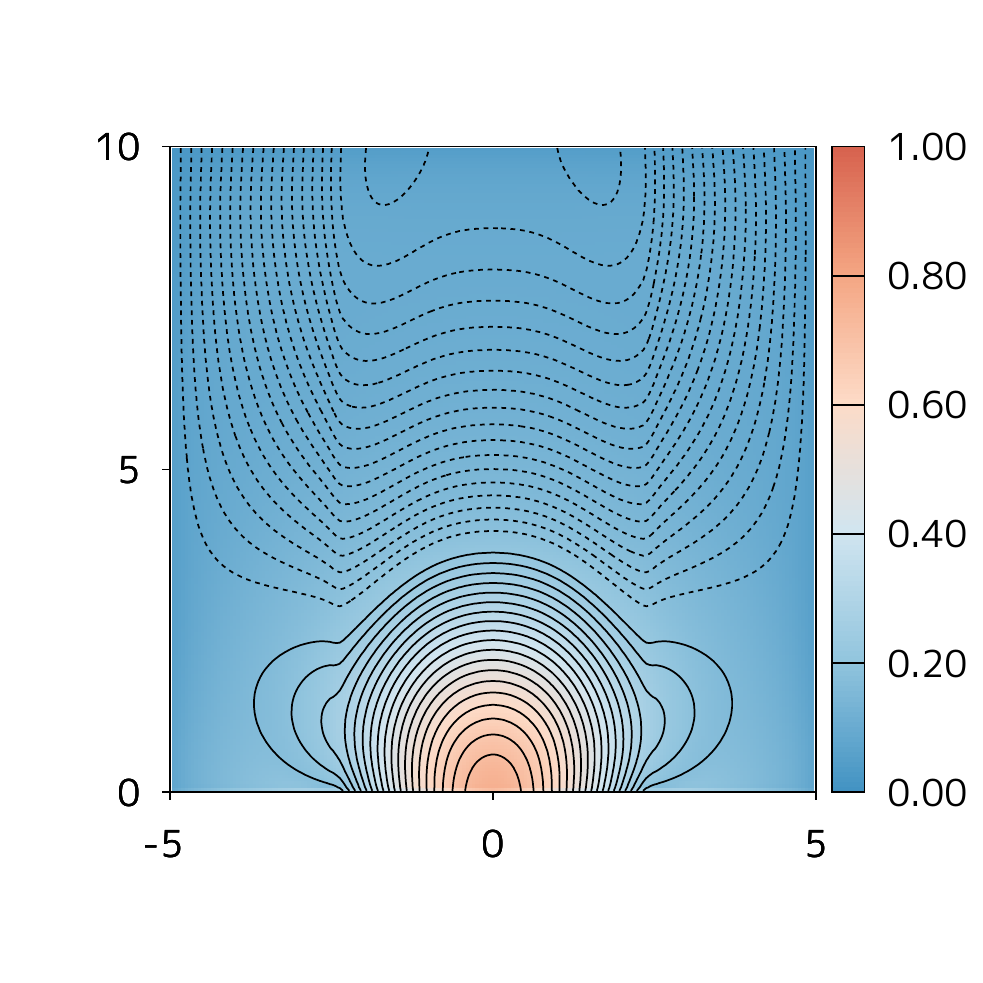}{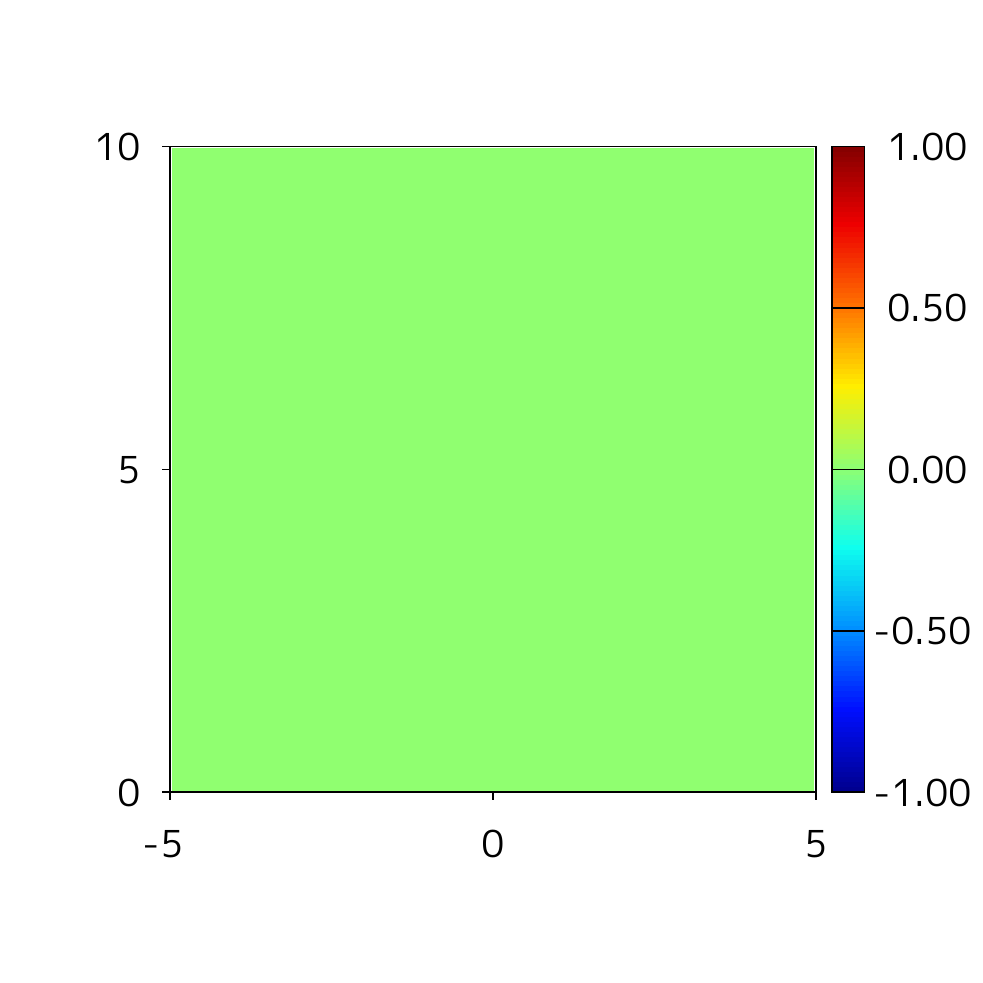}
\caption{
Same as Figure~\ref{fig:fig2}, but for the NLFFF.
\label{fig:fig3}}
\end{figure}

The present method results in the relaxation method for the NLFFF model
if forcing the pressure deviation to zero.
Thus, the NLFFF model is also computed for comparison.
The distribution of the magnetic field is shown in Figure~\ref{fig:fig3}.
The lower magnetic loop expands due to force imbalance at the bottom boundary,
while the upper loop seems to be suppressed by a detached magnetic field.
Here we notice that the detached magnetic field lines
seem to penetrate the upper boundary incorrectly.
It suggests that the NLFFF model with the NFFF boundary condition (\ref{eq:bessel})
does not satisfy the divergence-free condition with high accuracy.

\begin{figure}[t]
\centering
\plotone{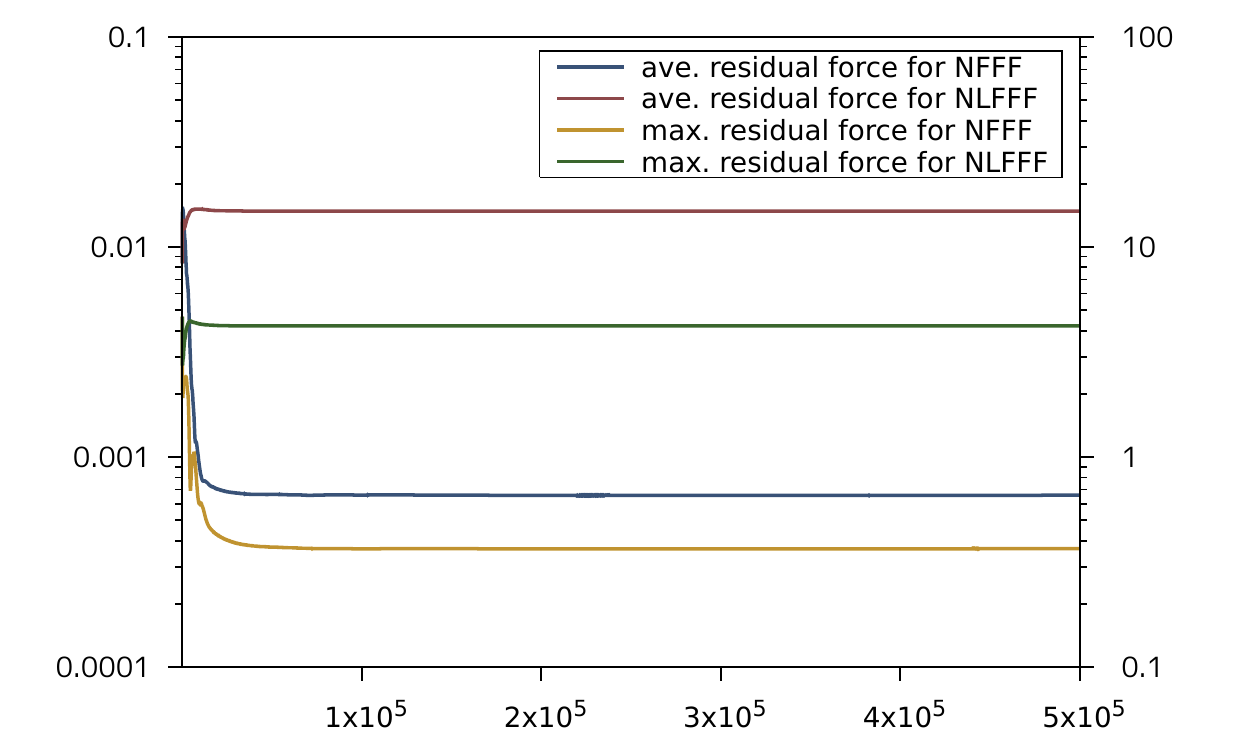}
\caption{
The residual force as a function of iteration number.
The left vertical axis represents the mean of the residual force,
and the right vertical axis represents the maximum value of the residual force.
The bluish, reddish, yellowish, and greenish lines correspond to the mean for the NFFF,
for the NLFFF, the maximum value for the NFFF, and for the NLFFF, respectively.
\label{fig:fig4}}
\end{figure}

Figure~\ref{fig:fig4} shows the residual force, i.e.,
the first and second terms of the right-hand side of (\ref{eq:vnfff}),
as a function of iteration number.
Both the mean and maximum values of the residual force for the NLFFF model
are larger than one order magnitude compared with those for the NFFF model.
The increase in the residual force for the NLFFF model is expected to be
due to the significant violation of the divergence-free condition near the bottom boundary.

\begin{figure}[t]
\centering
\plottwo{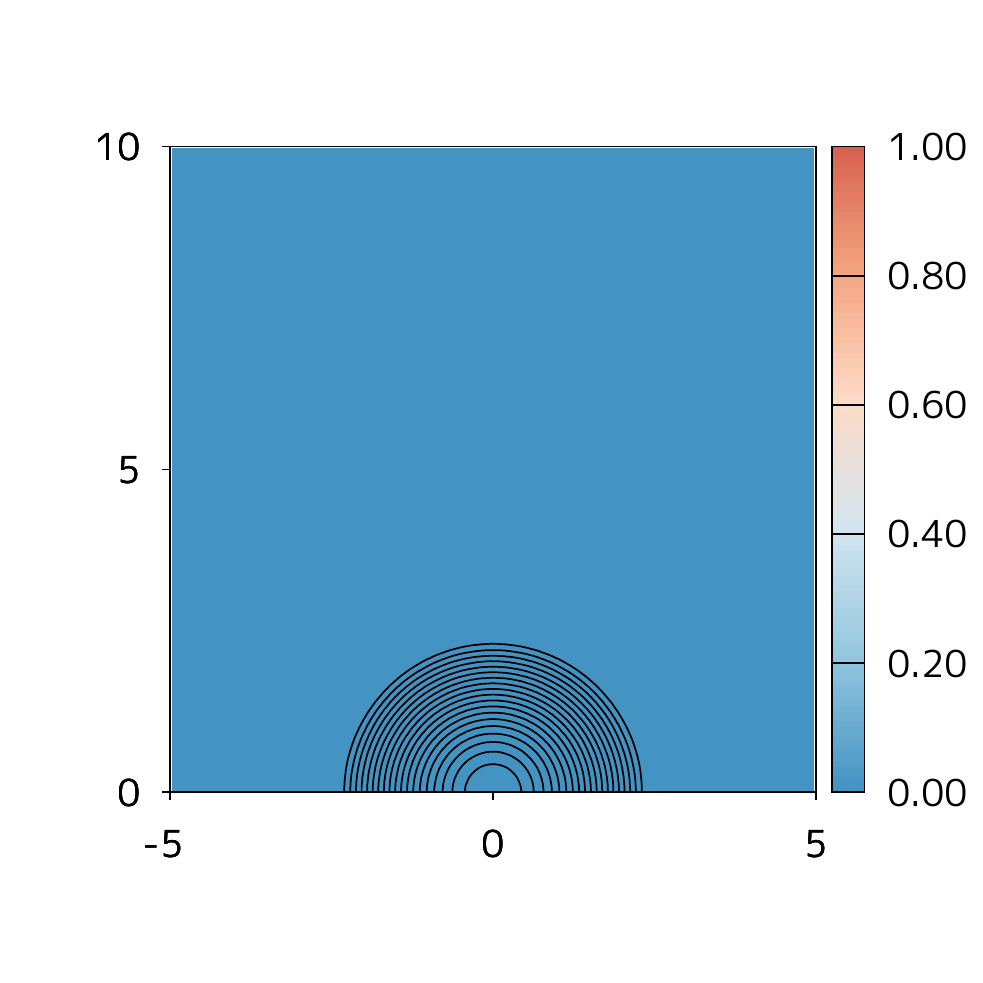}{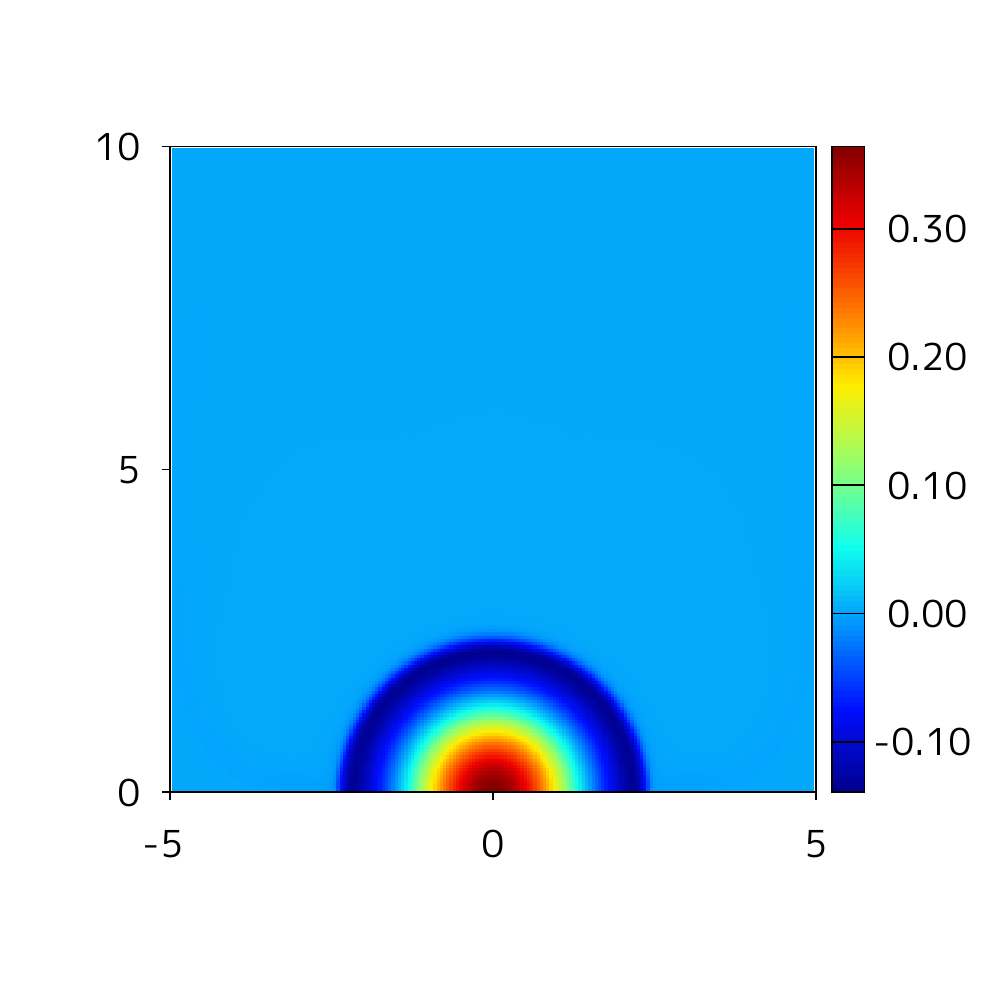}
\caption{
Same as Figure~\ref{fig:fig2}, but for $B_y = 0$ at the bottom boundary.
\label{fig:fig5}}
\end{figure}

We also perform the test without the out-of-plane magnetic field
$B_y$ at the bottom boundary instead of $B_y = J_0(x)$ in (\ref{eq:bessel}).
Since even the magnetic field at the boundary is completely
not force-free in this case,
an equilibrium distribution far away from force-free is expected.
Figure~\ref{fig:fig5} reveals that a semicircular magnetic field distribution
similar to the poloidal field distribution in Figure~\ref{fig:fig2}
is reconstructed, while the pressure deviation is automatically developed
so as to balance the inward Lorentz force
instead of the magnetic pressure due to $B_y$.
Conversely, the solution of the NLFFF model does not converge in this situation.

\subsection{Case $H(z) = Const.$}

\begin{figure}[t]
\centering
\plottwo{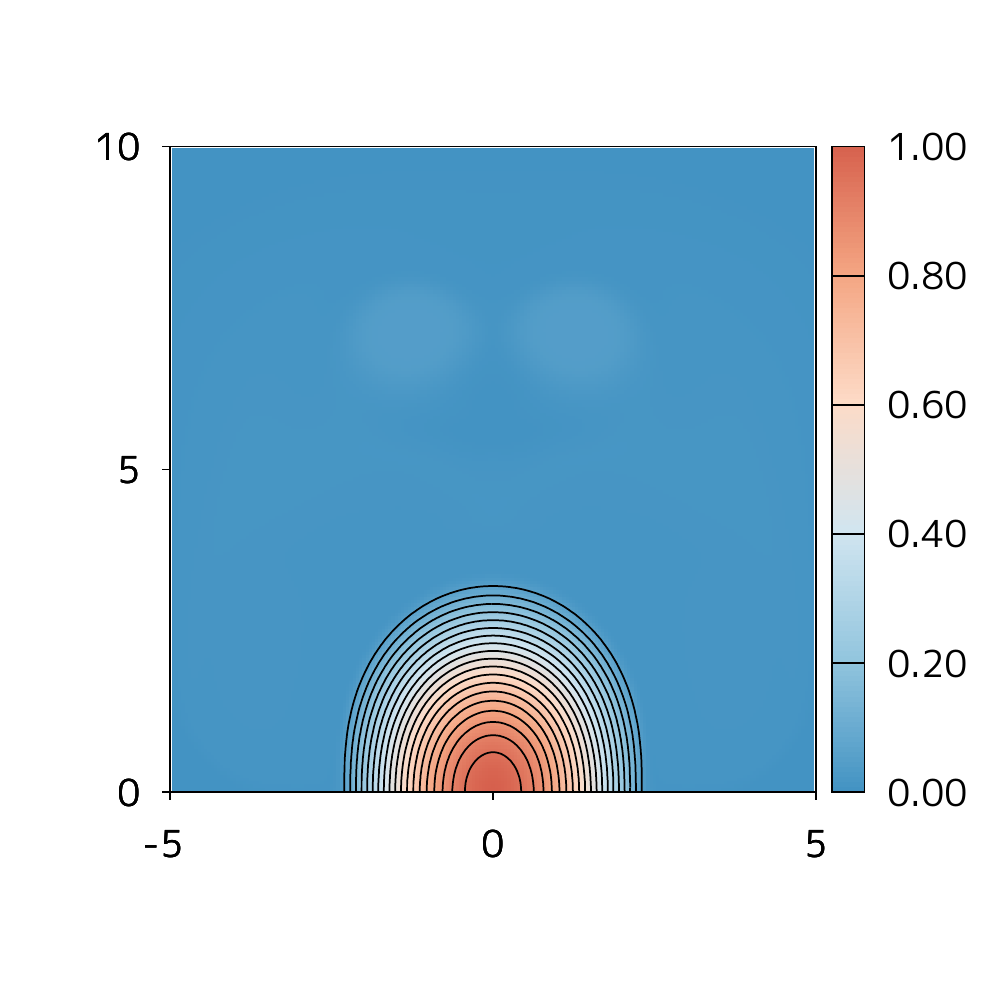}{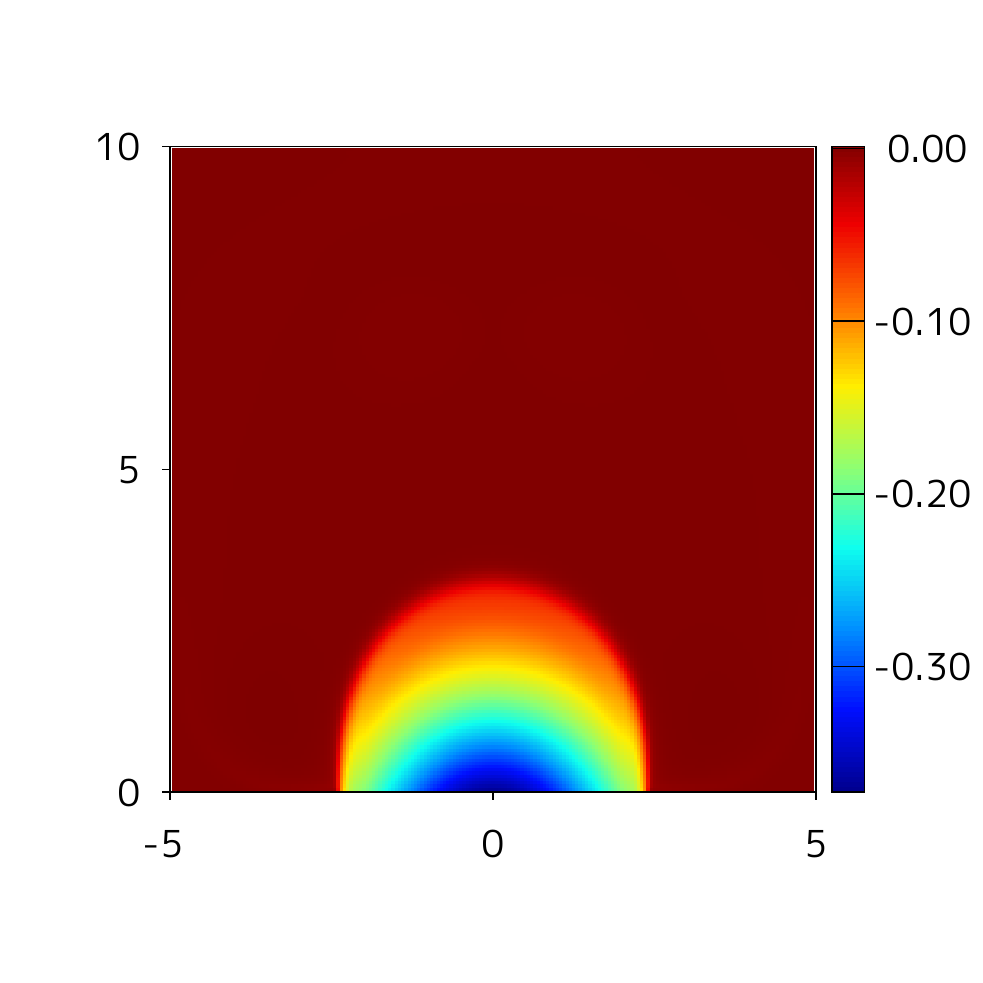}
\caption{
Same as Figure~\ref{fig:fig2}, but for $H(z) = 3$.
\label{fig:fig6}}
\end{figure}

When $H(z)$ is a sufficiently large constant as $100$,
a semi-cylindrical magnetic field distribution quite similar to that
in Figure~\ref{fig:fig2} is obtained.
However, if $H(z)$ becomes almost the same or smaller size as an active region,
the gravitational effect should be clearly seen.
Figure~\ref{fig:fig6} shows the results for the case $H(z) = 3$.
The gravitational stratification leads that
the pressure deviation decreases as the height increases.
Accordingly, the magnetic loop is elongated upward
because the pressure gradient force that suppress expansion
of the outer magnetic loop becomes weak.
Since the magnetic pressure much exceeds the pressure deviation at above the scale height,
the reconstructed magnetic field gradually approaches the force-free field.
Such configuration never be realized in the NLFFF model.

\subsection{Case $H(z) \neq Const.$}

Here we adopt the following scale height profile:
\begin{equation}
	H(z) = H_{ph} + \frac{1}{2} \left( H_{co} - H_{ph} \right)
	\left[ 1 + \tanh \left( \frac{z-Z_{ch}}{w} \right) \right],
	\label{eq:scalehs}
\end{equation}
where $H_{ph}$, $H_{co}$, $Z_{ch}$, and $w$ correspond to
the scale height on the photosphere, that in the corona,
the thickness of the chromosphere,
and the width of the transition layer from the chromosphere
to the corona, respectively.
The parameters are fixed as
$H_{ph} = 3$, $H_{co} = 100$, and $w = 0.2$.

\begin{figure}[t]
\centering
\plottwo{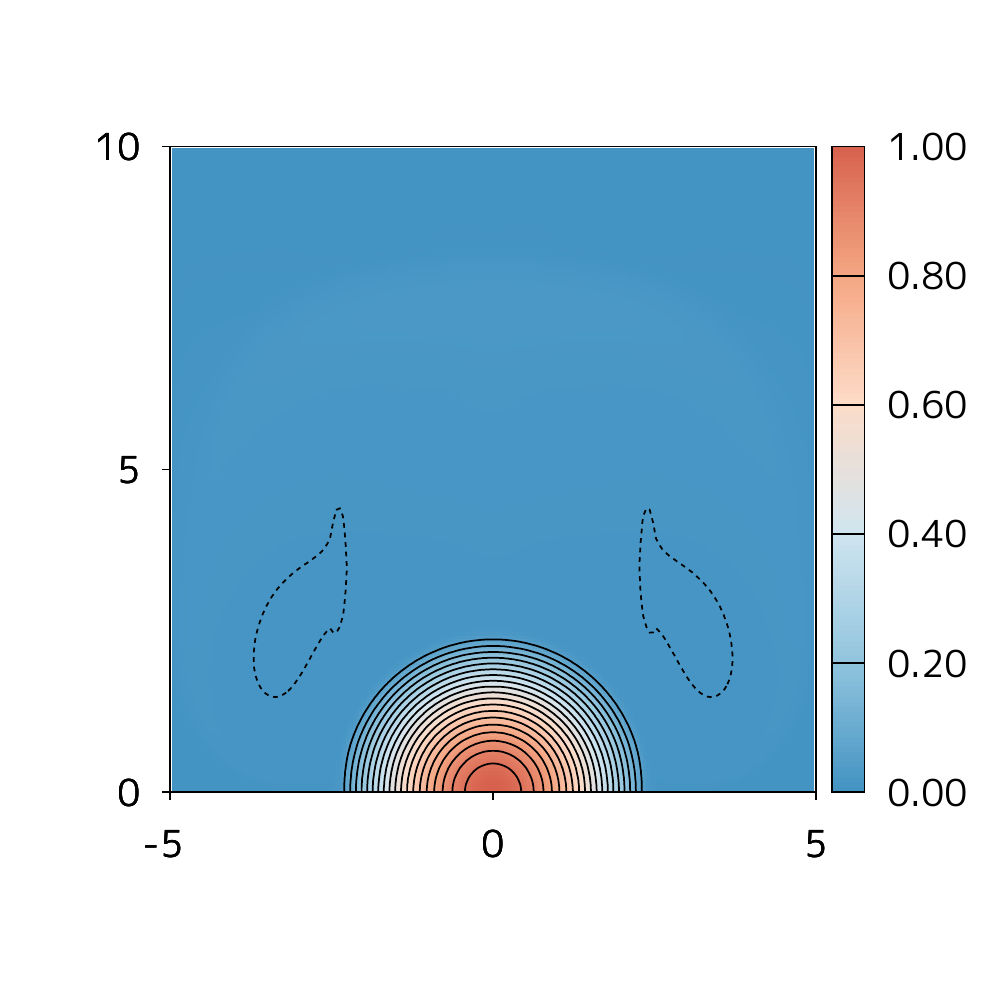}{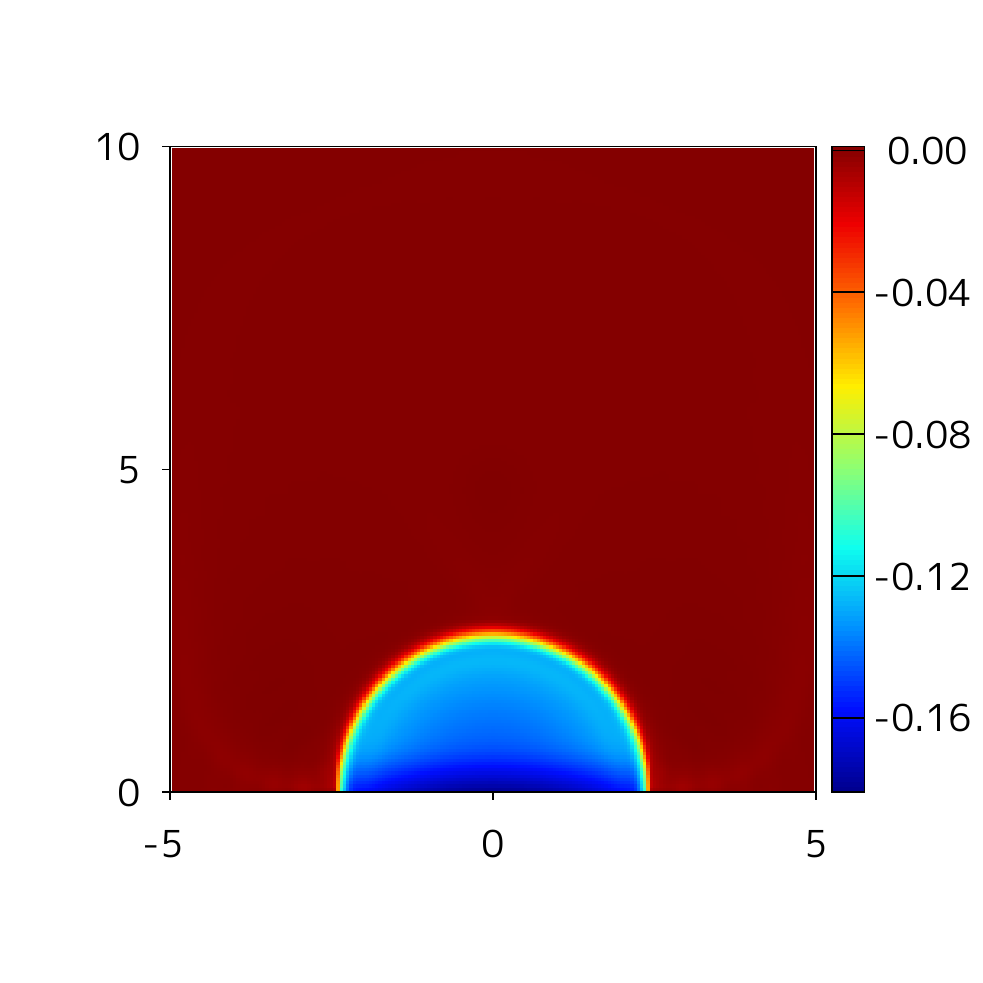}
\caption{
Same as Figure~\ref{fig:fig2},
but for that $H(z)$ is defined by (\ref{eq:scalehs}) with $Z_{ch} = 1$.
\label{fig:fig7}}
\end{figure}

\begin{figure}[t]
\centering
\plottwo{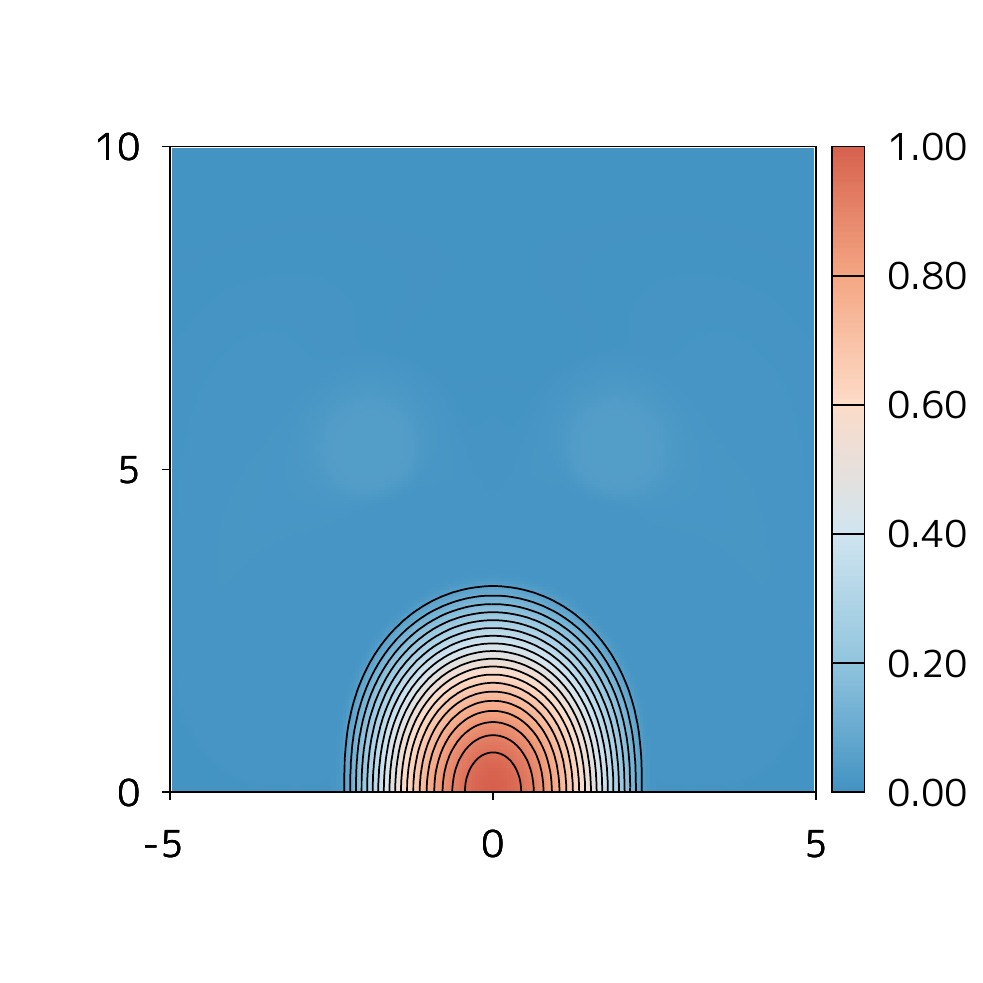}{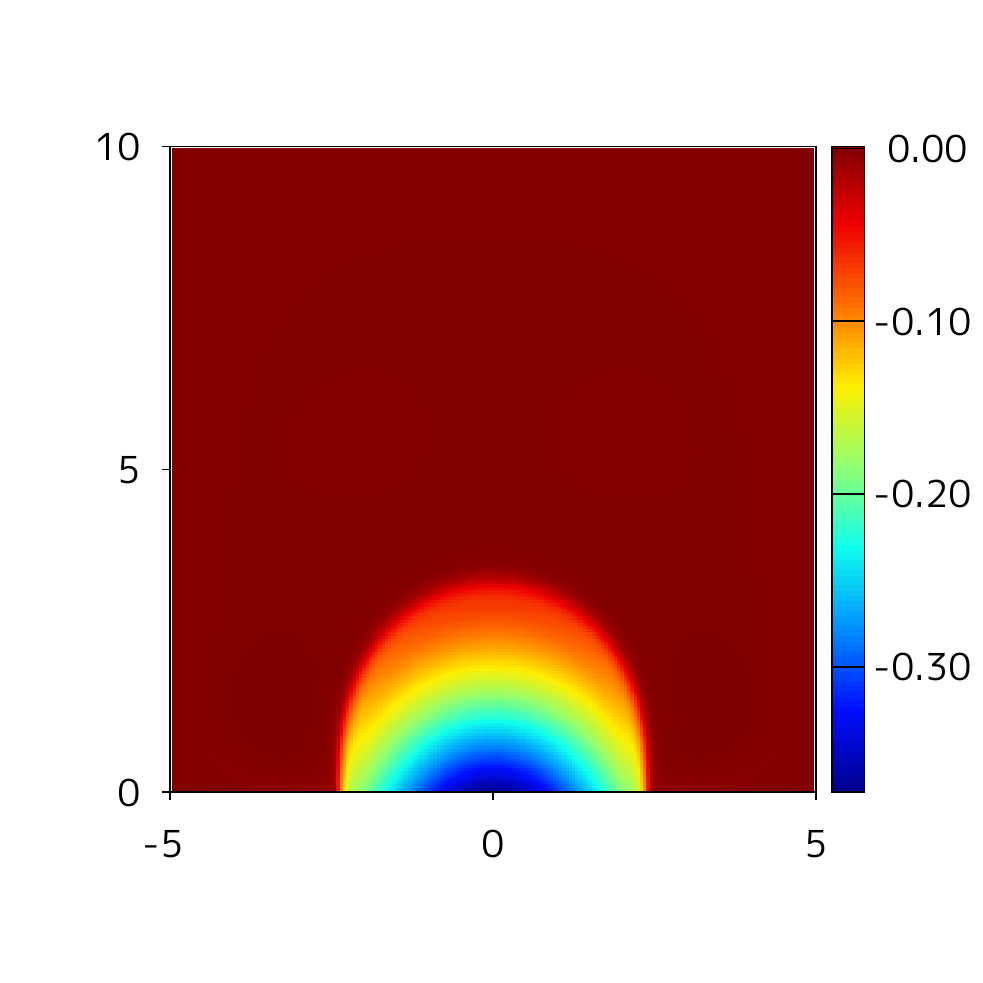}
\caption{
Same as Figure~\ref{fig:fig2},
but for that $H(z)$ is defined by (\ref{eq:scalehs}) with $Z_{ch} = 4$.
\label{fig:fig8}}
\end{figure}

Figure~\ref{fig:fig7} shows the results
for (\ref{eq:scalehs}) with $Z_{ch} = 1$.
The magnetic field distribution apparently resembles
that for Figure~\ref{fig:fig2}.
The pressure deviation decreases as the height increases by $z \approx 1$,
whereas the pressure deviation at $z > 1$, i.e., the corona in this model,
does not decrease significantly because $H(z)$ becomes large as $100$.
Therefore, the coronal magnetic field does not converge to the force-free field
since the magnetic field in the lower corona is not force-free.
On the other hand,
when $Z_{ch} = 4$ that is larger than the scale height in the chromosphere,
the pressure deviation decreases sufficiently in the chromosphere
as in Figure~\ref{fig:fig8}.
Thus, the distribution of the magnetic field as well as
the pressure deviation are similar to Figure~\ref{fig:fig6}.

The results of the present preliminary tests suggest that
small-scale magnetic fields are strongly affected
by the plasma pressure and/or the gravity.

\section{Summary} \label{sec:summary}

A new MHD relaxation method for the NFFF where the Lorentz force
balances the pressure gradient and gravitational forces
has been proposed in this paper.
In particular, assuming that a temperature profile in the solar atmosphere
is given beforehand and depends only on the height $z$,
the MHD equilibrium problem is reduced to solving
the magnetic field and pressure deviation for a given scale height $H(z)$.
Therefore, the basic equations (\ref{eq:vnfff}) to (\ref{eq:pnfff}) are proposed,
where the pseudo-speed of sound that is not physical but numerical is employed
in the evolution equation for the pressure deviation.
Note here that the pressure deviation can be negative
although the pressure should remain positive.
The backgroud pressure, which is determined independently of our model,
must be given so that the total plasma pressure is positive.
In the region of large $H(z)$,
the pressure deviation tends to be uniform along the magnetic field
according to the telegraph equation.
A sufficient condition for gravitational stability of the system
is also derived.

The basic equations are solved imposing only the vector magnetic field
on the photosphere,
where a robust high-resolution shock capturing scheme can be applied.
In order to confirm the effectiveness of our method
and show the difference from the NLFFF extrapolation,
we particularly performed two-dimensional numerical experiments.
The magnetic flux loops were able to be confined with a finite area
due to the pressure gradient force in the NFFF model
though those expanded to the whole domain in the NLFFF model.
Some demonstrations including the gravitational effect
were also carried out in this paper.

The critical assumption of the present model is that
the temperature distribution is given as a function only of $z$.
The realistic temperature distribution in the solar atmosphere,
however, is not plane parallel but more complicated.
When applying the present method to real data,
it is necessary to utilize a typical or an average vertical temperature profile
in the active region.
Moreover, since real data is dynamic and not completely MHS equilibrium,
some sort of preprocessing for the photospheric vector magnetic field
as in the NLFFF model \citep{wiegelmann06b,fuhrmann07} may be needed.
Thus, hereafter, we need further experiments that
three-dimensional magnetic fields are reconstructed using
the linear MHS model \citep{low91,zhu18} or observational dataset,
in which the temperature must be a three-dimensional distribution.
A comparative study with a time-dependent MHD simulation is also worthwhile.
We investigate, through the experiments, the applicability and extension
of the present method whose temperature depends only on $z$.
These are currently being studied and will be reported
in our future papers.

\acknowledgments

This work was supported by MEXT/JSPS KAKENHI Grant Numbers JP15K04756 (T.M.),
15H05812 and JP15H05814 (K.K.).

\appendix

\section{Stability Condition} \label{sec:disp}

Consider here a non-magnetic system, i.e., pure hydrodynamic system.
We assume periodic boundaries in all directions
and a constant scale height $H$ for simplicity.
Since the system becomes linear for the non-magnetic system,
the time evolution of Fourier modes as
\begin{equation}
	\left[
	\begin{array}{c}
		\hat{\vec{V}} \\
		\hat{\tilde{p}}
	\end{array}
	\right]
	\exp \left( - i \omega t + i \vec{k} \cdot \vec{r} \right),
	\label{eq:modes}
\end{equation}
where $\vec{k}$ is the wave vector and $\omega$ is the complex frequency,
can be exactly solved.
Substituting (\ref{eq:modes}) into (\ref{eq:vnfff}) with $\vec{B} = 0$
and (\ref{eq:pnfff}), we obtain the following homogeneous linear system:
\begin{equation}
	\left[
	\begin{array}{cccc}
		\omega + i \nu & 0 & 0 & - k_x \\
		0 & \omega + i \nu & 0 & - k_y \\
		0 & 0 & \omega + i \nu & - k_z + i/H \\
		- a^2 k_x & - a^2 k_y & - a^2 k_z & \omega
	\end{array}
	\right]
	\left[
	\begin{array}{c}
		\hat{V_x} \\
		\hat{V_y} \\
		\hat{V_z} \\
		\hat{\tilde{p}}
	\end{array}
	\right]
	= 0.
	\label{eq:homo}
\end{equation}
This system has nontrivial solutions
if the determinant of the coefficient matrix is zero.
Thus, we obtain the dispersion relation (\ref{eq:disp}) and rewrite it as
\begin{equation}
	\left( \omega + i \nu \right)^2
	\left( \omega^2 + i \nu \omega - a^2 k^2 + i \frac{a^2 k_z}{H} \right)
	= 0.
	\label{eq:disp2}
\end{equation}
The solutions of (\ref{eq:disp2}) are
\begin{equation}
	\omega = - i \nu,
	\label{eq:om1}
\end{equation}
and
\begin{equation}
	\omega = - i \frac{\nu}{2} \pm \sqrt{\xi},
	\label{eq:om2}
\end{equation}
where
\begin{equation}
	\xi = a^2 k^2 - \frac{\nu^2}{4} - i \frac{a^2 k_z}{H}.
	\label{eq:om2xi}
\end{equation}

The stability of the system is determined by the imaginary part of $\omega$.
The amplitudes of the Fourier modes (\ref{eq:modes}) exponentially grow in time
if $Im(\omega) > 0$, while those exponentially decay in time if $Im(\omega) < 0$.
Since $Im(\omega)$ of (\ref{eq:om1}) is always negative for $\nu > 0$,
the modes are stable.
On the other hand, $Im(\omega)$ of (\ref{eq:om2}) as
\begin{equation}
	Im(\omega) = - \frac{\nu}{2} \pm \sqrt{\frac{\left| \xi \right|-Re(\xi)}{2}}
	\label{eq:om2im}
\end{equation}
are always negative when the condition as
\begin{equation}
	\frac{\nu^2}{2} > \left| \xi \right| - Re(\xi)
	\label{eq:om2imp}
\end{equation}
is satisfied. The stability condition (\ref{eq:om2imp}) leads to
\begin{equation}
	\nu > \frac{a}{H} \frac{\left| k_z \right|}{\left| k \right|}
	\label{eq:om2imp2}
\end{equation}
after some manipulations.
Thus, the sufficient condition for the stability of the system is given by
\begin{equation}
	\nu > \frac{a}{H}
	\label{eq:om2imp3}
\end{equation}
for any wave vectors.
The system is expected to be more stable when wall boundaries are considered.

\section{Eigenvalues} \label{sec:eigen}

Consider the one-dimensional system of
the Eqs. (\ref{eq:cons}) without the source terms as
\begin{equation}
	\frac{\partial}{\partial t}
	\left[
	\begin{array}{c}
		V_x \\
		V_y \\
		V_z \\
		B_y \\
		B_z \\
		\tilde{p}
	\end{array}
	\right]
	+
	\frac{\partial}{\partial x}
	\left[
	\begin{array}{c}
		\tilde{p} + \frac{B_y^2}{2} + \frac{B_z^2}{2} - \frac{B_x^2}{2} \\
		- B_x B_y \\
		- B_x B_z \\
		V_x B_y - B_x V_y \\
		V_x B_z - B_x V_z \\
		a^2 V_x
	\end{array}
	\right]
	= 0,
	\label{eq:cons1d}
\end{equation}
where $B_x$ is constant due to the divergence-free condition
for the magnetic field.
The Jacobian $A$ of (\ref{eq:cons1d}) is given by
\begin{equation}
	A =
	\left[
	\begin{array}{ccccccc}
		0 & 0 & 0 & B_y & B_z & 1 \\
		0 & 0 & 0 & -B_x & 0 & 0 \\
		0 & 0 & 0 & 0 & -B_x & 0 \\
		B_y & -B_x & 0 & V_x & 0 & 0 \\
		B_z & 0 & -B_x & 0 & V_x & 0 \\
		a^2 & 0 & 0& 0 & 0 & 0
	\end{array}
	\right].
	\label{eq:jacob}
\end{equation}
The eigenvalues $\lambda$ of $A$ satisfy the following characteristic equation:
\begin{equation}
	\left\{ \lambda \left( \lambda - V_x \right) - B_x^2 \right\}
	\left\{ \lambda^4 - V_x \lambda^3
	- \left( B^2 + a^2 \right) \lambda^2
	+ V_x a^2 \lambda + B_x^2 a^2 \right\} = 0.
	\label{eq:eigen}
\end{equation}
where $B^2 = B_x^2+B_y^2+B_z^2$.
Two of the roots of (\ref{eq:eigen}) are
\begin{equation}
	\lambda = \frac{V_x \pm \sqrt{V_x^2+4 B_x^2}}{2}.
	\label{eq:eigen1}
\end{equation}
These correspond to the Alfv\'{e}n waves.
The other four roots are obtained from
\begin{equation}
	\lambda^4 - V_x \lambda^3
	- \left( B^2 + a^2 \right) \lambda^2
	+ V_x a^2 \lambda + B_x^2 a^2 = 0.
	\label{eq:eigen4eq}
\end{equation}
If $V_x = 0$, then (\ref{eq:eigen4eq}) gives
\begin{equation}
	\lambda = \pm \sqrt{\frac{B^2+a^2 \pm
	\sqrt{\left( B^2+a^2 \right)^2 - 4 B_x^4 a^4}
	}{2}}.
	\label{eq:eigen2}
\end{equation}
Moreover, if $B_y = B_z = 0$,
\begin{equation}
	\lambda = \frac{V_x \pm \sqrt{V_x^2+4 B_x^2}}{2} \ , \ \
	\lambda = \pm a.
	\label{eq:eigen3}
\end{equation}
Therefore, the roots of (\ref{eq:eigen4eq}) indicate the magnetosonic waves.
Since the eigenvalues are real, the system (\ref{eq:cons1d}) is hyperbolic.

\section{Upwind-type Scheme} \label{sec:upwind}

Consider first the one-dimensional system of (\ref{eq:cons1d})
as in Appendix \ref{sec:eigen}. Since the system is hyperbolic,
an upwind-type numerical scheme enables us to realize robust computation.
However, construction of Godunov's scheme or Roe's scheme for the system
is neither so easy nor necessary.
Therefore, a simple numerical scheme is developed here,
where numerical fluxes at a cell interface are evaluated using
approximate systems of equations instead of (\ref{eq:cons1d}).
We split the system of (\ref{eq:cons1d}) into compressible
($\partial V_x / \partial x \neq 0$) and incompressible
($\partial V_x / \partial x = 0$) systems as follows:
The compressible system is given by
\begin{equation}
	\frac{\partial}{\partial t} \left[
	\begin{array}{c}
		V_x \\
		\tilde{p}{{}_T}_\perp
	\end{array}
	\right] + \left[
	\begin{array}{cc}
		0 & 1 \\
		a^2+B_\perp^2 & 0
	\end{array}
	\right] \frac{\partial}{\partial x} \left[
	\begin{array}{c}
		V_x \\
		\tilde{p}{{}_T}_\perp
	\end{array}
	\right] = 0,
	\label{eq:fvxp}
\end{equation}
where $B_\perp^2 = B_y^2+B_z^2$ and $\tilde{p}{{}_T}_\perp = \tilde{p} + B_\perp^2 / 2$.
The incompressible systems are given by
\begin{equation}
	\frac{\partial}{\partial t} \left[
	\begin{array}{c}
		V_y \\
		B_y
	\end{array}
	\right] + \left[
	\begin{array}{cc}
		0 & - B_x \\
		-B_x & V_x
	\end{array}
	\right] \frac{\partial}{\partial x} \left[
	\begin{array}{c}
		V_y \\
		B_y
	\end{array}
	\right] = 0,
	\label{eq:fvyby}
\end{equation}
\begin{equation}
	\frac{\partial}{\partial t} \left[
	\begin{array}{c}
		V_z \\
		B_z
	\end{array}
	\right] + \left[
	\begin{array}{cc}
		0 & - B_x \\
		-B_x & V_x
	\end{array}
	\right] \frac{\partial}{\partial x} \left[
	\begin{array}{c}
		V_z \\
		B_z
	\end{array}
	\right] = 0,
	\label{eq:fvzbz}
\end{equation}
respectively.
Thus, the magnetosonic waves and the planar Alfv\'{e}n waves are decoupled.
We readily derive the upwind-biased values at the cell interface, denoted by $h$,
from (\ref{eq:fvxp}) as
\begin{equation}
	V_x^h = \frac{V_x^R + V_x^L}{2} - \frac{1}{2 \sqrt{a^2+B_\perp^2}}
	\left( \tilde{p}{{}_T}_\perp^R - \tilde{p}{{}_T}_\perp^L \right),
	\label{eq:vxh}
\end{equation}
\begin{equation}
	\tilde{p}{{}_T}_\perp^h = \frac{\tilde{p}{{}_T}_\perp^R + \tilde{p}{{}_T}_\perp^L}{2}
	- \frac{\sqrt{a^2+B_\perp^2}}{2} \left( V_x^R - V_x^L \right),
	\label{eq:ph}
\end{equation}
where the superscripts $R$ and $L$ indicate the right and left states at the interface.
Here $B_\perp^2$ is given by $\max \left( {B_\perp^R}^2 , {B_\perp^L}^2 \right)$
because $B_\perp$ is not determined self-consistently.
From (\ref{eq:fvyby}),
\begin{equation}
	V_y^h = \frac{V_y^R + V_y^L}{2}
	+ \frac{V_x}{2 \sqrt{V_x^2+4 B_x^2}} \left( V_y^R - V_y^L \right)
	+ \frac{B_x}{\sqrt{V_x^2+4 B_x^2}} \left( B_y^R - B_y^L \right),
	\label{eq:vyh}
\end{equation}
\begin{equation}
	B_y^h = \frac{B_y^R + B_y^L}{2}
	+ \frac{B_x}{\sqrt{V_x^2+4 B_x^2}} \left( V_y^R - V_y^L \right)
	- \frac{V_x}{2 \sqrt{V_x^2+4 B_x^2}} \left( B_y^R - B_y^L \right),
	\label{eq:byh}
\end{equation}
and from (\ref{eq:fvzbz}),
\begin{equation}
	V_z^h = \frac{V_z^R + V_z^L}{2}
	+ \frac{V_x}{2 \sqrt{V_x^2+4 B_x^2}} \left( V_z^R - V_z^L \right)
	+ \frac{B_x}{\sqrt{V_x^2+4 B_x^2}} \left( B_z^R - B_z^L \right),
	\label{eq:vzh}
\end{equation}
\begin{equation}
	B_z^h = \frac{B_z^R + B_z^L}{2}
	+ \frac{B_x}{\sqrt{V_x^2+4 B_x^2}} \left( V_z^R - V_z^L \right)
	- \frac{V_x}{2 \sqrt{V_x^2+4 B_x^2}} \left( B_z^R - B_z^L \right)
	\label{eq:bzh}
\end{equation}
are obtained using the method of characteristics.
Here $V_x$ is replaced by (\ref{eq:vxh}).
Hence, using (\ref{eq:vxh}) to (\ref{eq:bzh}),
the numerical fluxes of (\ref{eq:cons1d}) can be evaluated as
\begin{equation}
	F^h = \left[
	\begin{array}{c}
		\tilde{p}{{}_T}_\perp^h - \frac{B_x^2}{2} \\
		- B_x B_y^h \\
		- B_x B_z^h \\
		V_x^h B_y^h - B_x V_y^h \\
		V_x^h B_z^h - B_x V_z^h \\
		a^2 V_x^h
	\end{array}
	\right].
	\label{eq:flux1d}
\end{equation}

In order to extend to multidimensions,
the divergence-free constraint of the magnetic field must be maintained numerically.
Although variants of Constrained-Transport (CT) method \citep[e.g.,][]{toth00,minoshima19}
where the divergence-free constraint is exactly preserved can be applied,
here we adopt the hyperbolic divergence cleaning method \citep{dedner02}.
Introduce a new scalar field $\psi$ that corrects the magnetic field,
and consider the time evolution of the numerical divergence of the magnetic field.
In order to reduce the divergence of the magnetic field $\partial B_x / \partial x$,
\begin{equation}
	\frac{\partial}{\partial t} \left[
	\begin{array}{c}
		B_x \\
		\psi
	\end{array}
	\right] + \left[
	\begin{array}{cc}
		0 & 1 \\
		c_h^2 & 0
	\end{array}
	\right] \frac{\partial}{\partial x} \left[
	\begin{array}{c}
		B_x \\
		\psi
	\end{array}
	\right] = 0
\end{equation}
are solved in addition to (\ref{eq:cons1d}).
The upwind-biased values at the interface are evaluated as
\begin{equation}
	B_x^h = \frac{B_x^R + B_x^L}{2}
	- \frac{1}{2 c_h} \left( \psi^R - \psi^L \right),
	\label{eq:bxh}
\end{equation}
\begin{equation}
	\psi^h = \frac{\psi^R + \psi^L}{2}
	- \frac{c_h}{2} \left( B_x^R - B_x^L \right).
	\label{eq:psih}
\end{equation}
Constant $B_x$ in the numerical fluxes of (\ref{eq:flux1d}) is replaced
by (\ref{eq:bxh}).

\section{Boundary Conditions} \label{sec:boundary}

At the bottom boundary that corresponds to the photosphere,
the following conditions in the ghost cell are applied:
\begin{equation}
	\left\{
	\begin{array}{l}
		\vec{B}_G = \vec{B}_{ph}, \\
		\vec{V}_G = 0, \\
		\tilde{p}_G = \tilde{p}_I - {V_z}_I \sqrt{a^2+B_{ph}^2}, \\
		\psi_G = 0,
	\end{array}
	\right.
	\label{eq:bcb}
\end{equation}
where the subscripts $G$ and $I$ are the indices corresponding to
the ghost cell and the cell just inside the boundary, respectively.
The unphysical scalar field $\psi$ correcting the magnetic field
is fixed in time because of numerical stability.

At the top and lateral boundaries,
\begin{equation}
	\left\{
	\begin{array}{l}
		\vec{B}_G = 0, \\
		{V_\mathrm{n}}_G = {V_\mathrm{n}}_I + \tilde{p}_I / a , \\
		{\vec{V}_\mathrm{t}}_G = 0, \\
		\tilde{p}_G = 0, \\
		\psi_G = \psi_I + c_h {B_\mathrm{n}}_I
	\end{array}
	\right.
	\label{eq:bco}
\end{equation}
are imposed in the numerical experiments.
Here $\mathrm{n}$ and $\mathrm{t}$ indicate the normal and tangential components,
in which the normal vector points outward from the domain.
Assuming that $\tilde{p}_G$ and an approximate characteristics as
$dp+dV_{\mathrm{n}}a$ are zero, ${V_\mathrm{n}}_G$ is calculated.
Unlike at the bottom boundary, $\psi_G$ is extrapolated from the inner quantities
though the condition for $\psi$ does not much affect the results.





\allauthors

\listofchanges

\end{document}